# Micro-transfer-printed Thin film lithium niobate (TFLN)-on-Silicon Ring Modulator

YING TAN,[1, 2, *] SHENGPU NIU,[2, 3] MAXIMILIEN BILLET,[1,2,4] NISHANT SINGH,[2, 3] MARGOT NIELS,[1,2] TOM VANACKERE,[1,2,4] JORIS VAN KERREBROUCK,[2,3] GUNTHER ROELKENS,[1,2] BART KUYKEN,[1,2] AND DRIES VAN THOURHOUT[1, 2, *]

[1] *Photonics Research Group, Department of information Technology (INTEC), Ghent University–imec, 9052, Ghent, Belgium*
[2] *IMEC, Kapeldreef 75, 3001 Heverlee, Belgium*
[3] *IDLab, Department of information Technology (INTEC), Ghent University–imec, 9052, Ghent, Belgium*
[4] *OPERA-Photonique CP 194/5, Universitie Libre de Bruxelles (ULB), Bruxelles, Belgium*
*ying.tan@ugent.be, dries.vanthourhout@ugent.be*

**Abstract:** Thin-film lithium niobate (TFLN) has a proven record of building high-performance electro-optical (EO) modulators. However, its CMOS incompatibility and the need for non-standard etching have consistently posed challenges in terms of scalability, standardization, and the complexity of integration. Heterogeneous integration comes to solve this key challenge. Micro-transfer printing of thin-film lithium niobate brings TFLN to well-established silicon ecosystem by easy "pick and place", which showcases immense potential in constructing high-density, cost-effective, highly versatile heterogeneous integrated circuits. Here, we demonstrated for the first time a micro-transfer-printed thin film lithium niobate (TFLN)-on-silicon ring modulator, which is an important step towards dense integration of performant lithium niobate modulators with compact and scalable silicon circuity. The presented device exhibits an insertion loss of−1.5dB, extinction ratio of−37dB, electro-optical bandwidth of 16GHz and modulation rates up to 45Gbits$^{-1}$.

## 1. Introduction

The lithium niobate (LN) modulator is renowned for its exceptional performance and has a well-established history of commercial use [1–5]. Thin-film lithium niobate modulators have been demonstrated with record-breaking performance, with over 100Gbits$^{-1}$ modulation speed and <1V driving voltages[6–8]. The key limitation of LN is its CMOS incompatibility, which poses significant challenges in terms of scalability, production cost, standardization, versatility and integration complexity, as shown in Table 1. As such, there is a strong demand for heterogeneous integration of TFLN. Micro-transfer printing (µ-TP) of TFLN [9,10] offers a low-cost scalable heterogeneous solution with high integration density and reduced fabrication complexity.

Lithium niobate possesses a strong Pockels effect, which enables low-loss ultrafast EO modulation. The emergence of thin-film lithium niobate and the development of low-loss etching processes greatly enhanced its modulation efficiency[11–14]. High performance TFLN modulators have been demonstrated with low-loss, high speed and low drive voltages and high signal fidelity[6–8,15,16]. However, the lithium niobate on insulator (LNOI) platform has its limitations. Firstly, LNOI has low scalability because of its limited wafer size and its relatively small index contrast. Secondly, LNOI is not CMOS compatible due to the risk for lithium contamination. In addition, LN is difficult to etch and it usually requires non-standard etching techniques. The resulting ridge waveguides with sloped side-walls also make standardization difficult and limits the minimal feature size. The platform also lacks efficient grating couplers. Edge couplers[17–19] are commonly used on LNOI platforms to achieve high coupling efficiency, but they are sensitive to optical alignment and pose challenges for packaging.

In contrast, the silicon on insulator (SOI) platform stands as a leading photonic integration platform, characterized by high-index contrast and CMOS compatibility. High-index contrast enables compact devices, dense photonic routing and large-scale integration. CMOS compatibility promises low-cost and high-volume production. The extensive library of mature functional components further strengthens its appeal. However, modulation schemes on silicon predominantly rely on the plasma dispersion effect, leading to issues like AM/FM coupling and relatively high loss. Heterogeneous integration[20] incorporates TFLN into the well-established silicon ecosystem, expanding the range of modulation mechanisms available in the silicon platform and the range of application cases of thin-film lithium niobate.

Table 1 Comparison of different material platforms.

| Platform | CMOS compatible | $\chi^{(2)}$ Nonlinearity | Integration Density | Integration Complexity | Versatility | Scalability |
|---|---|---|---|---|---|---|
| *SOI* | ✓ | ✗ | high | low | high | high |
| *SiN* | ✓ | ✗ | low | low | low | high |
| *LNOI* | ✗ | ✓ | low | high | low | low |

Heterogeneous integration of TFLN can be realized by wafer bonding[21–24] or the newly emerging micro-transfer printing technology[9,25,26]. Micro-transfer printing of TFLN outperforms wafer bonding in fabrication complexity, integration density and cost-effectiveness. The fabrication process of µ-TP involves a source wafer and a target wafer. Suspended "coupons" are picked up using an elastomeric stamp and printed over the target wafer. These coupons are thin-film materials and thin-film devices, which do not include the substrate. This greatly simplifies the fabrication process by confining all the intricate fabrication procedures to the source wafer and minimizing the processing on the target chip. The preparation of the target chip on the other hand follows standard CMOS-compatible fabrication processes. In addition, micro-transfer printing enables printing miniaturized coupons over a very compact area, resulting in increased integration density on the target chip. Micro-transfer printing makes efficient use of the source material by pick up coupons of required material size rather than etching away excess material[27–30].

Micro-ring resonators[31–34] offer efficient modulation in a compact footprint, which facilitates dense integration with other functional elements. Micro-ring resonators are also compatible with wavelength-division-multiplexing (WDM), a widely applied and scalable approach for reaching higher data rates. While Mach-Zehnder interferometers are widely used for realizing LN modulators, they often have a large foot-print (~a few centimeters long), which is inconvenient for large-scale integration and packaging. In this work, we experimentally demonstrated micro-transfer-printed thin film lithium niobate (TFLN)-on-silicon ring modulator. The device has an insertion loss of ∼−1.5 dB and extinction ratio of −37dB. We also performed high-speed characterization of the modulator. The measured device exhibits an electro-optical bandwidth of 16 GHz and supports data rates up to 45Gbits$^{-1}$.

## 2. Design and fabrication

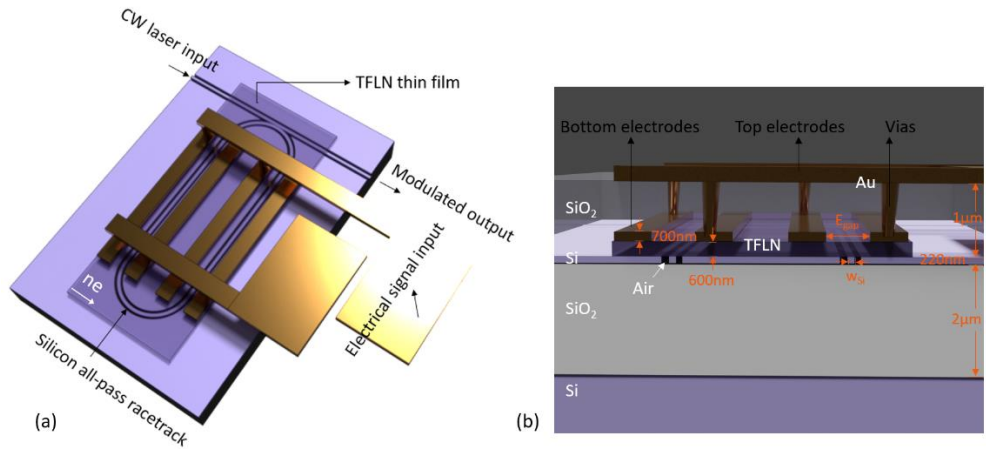

Fig. 1 Schematic image of the micro-transfer printed thin film lithium niobate (TFLN)-on-silicon ring modulator(top-view); (b) cross-sectional image of the modulator.

Fig. 1 (a) illustrates the modulator concept. The modulator is based on a hybrid Si-LN ring modulator, which consists of an all-pass racetrack resonator patterned in the silicon layer and an x-cut TFLN thin film over the top of the racetrack. By applying an electric field to the straight sections of the hybrid racetrack, the Pockels nonlinear effect in LN induces a linear electro-optic response, resulting a frequency shift of the resonator. Electric fields are applied in the same direction so that the phase change in both phase-shifters of the racetrack adds up. Fig. 1(b) shows a schematic image of the cross-section of the ring modulator. It includes a 220 nm thick silicon layer, a 600 nm thick TFLN layer and a 1μm thick silicon dioxide cladding layer. Two metal layers and vias were fabricated to enable convenient RF probing.

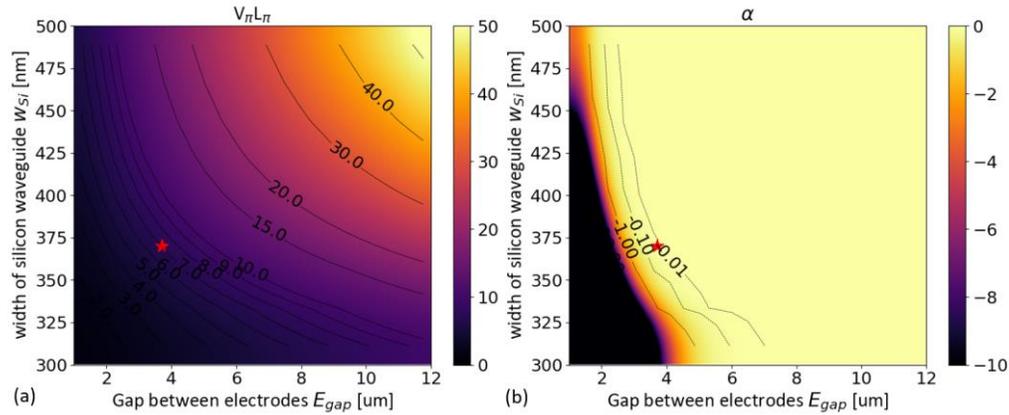

Fig. 2 Numerical optimization of micro-transfer printed thin film lithium niobate (TFLN)-on-Silicon ring modulator. (a) Simulation of the half-wave voltage-length product $V_\pi L_\pi$; (b) simulation of metal induced propagation loss of the hybrid waveguide.

The racetrack geometry and the Si-LN hybrid waveguide were optimized using numerical simulations through Comsol Multiphysics. The tuning efficiency of the micro-transfer printed thin film lithium niobate (TFLN)-on-Silicon ring modulator is mainly determined by 2 factors, the change of refractive index per volts $\Delta n_{eff,\ 1V}$ and the proportion of the ring covered with phase-shifters $f$, which is specified in Equation (1) and Equation (2). $\Delta n_{eff,\ 1V}$ is maximized when the applied electric field is aligned with the extraordinary axis of the x-cut lithium niobate thin film and the overlap of electric field and the optical field is maximized. Electro-optical tuning can be less efficient in the bending region. As bending is unavoidable in a ring configuration, the racetrack was designed with a long straight section *l*=400μm and very sharp bending *R*=15 μm. In this way, a large proportion of the ring is covered by phase-shifters *f* is 0.894. To enable sharp bending, we chose a waveguide width $w_{Si}$ of 370nm. In this case, a spacing of 3.7 μm is chosen, so that the metallic loss is maintained around 0.01dB/cm, as shown in Fig. 2(b). The $V_\pi L_\pi$ under the chosen parameter is 7V·cm, as shown in Fig. 2(a). $\Delta n_{eff,\,1V}$ is 1.1e−05 per volt and the tuning efficiency is 3.81pm/V, as calculated in Equation (2).

$$\Delta n_{eff,1V} \approx -\frac{c\varepsilon_0 n_{LN}^4 r_{33} V_0 \iint_{LN} \frac{E_{DC,y(x,y)}}{V_0} |E_x(x,y)|^2 \, dxdy}{2\iint_{-\infty}^{+\infty} \mathbb{R}\{E(x,y) \times H^*(x,y)\} \cdot e_z \, dxdy} \quad (1)$$

$$\frac{\delta\lambda_{res}}{\delta V} = f \times \frac{\lambda_{res}}{n_g} \times \Delta n_{eff,1V} \approx 0.895 \times \frac{1550nm}{4} \times 1.1e-05V^{-1} \approx 3.815 pmV^{-1} \quad (2)$$

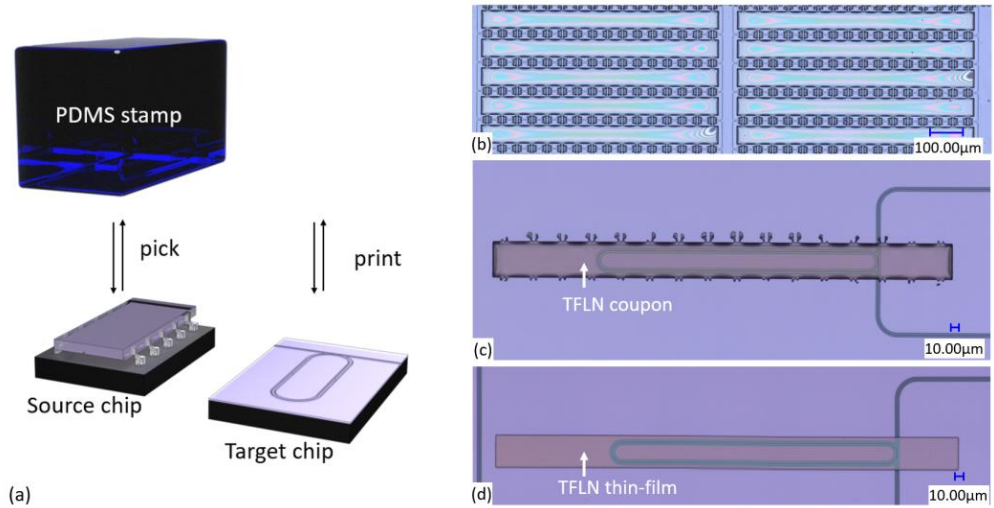

Fig. 3 (a) Illustrations of the transfer printing of thin-film lithium niobate to silicon circuits; (b) microscopic image of suspended coupons on the source chip; (c) microscope image of a silicon ring resonator with a printed thin-film lithium niobate coupon; (d) microscope image of a silicon ring resonator with a TFLN thin-film.

Fig. 3 (a) illustrates the micro-transfer printing process. Micro-transfer printing involves a source wafer and a target wafer. The target wafer contains silicon waveguide structures prepared by standard silicon processing. The source wafer contains suspended "TFLN coupon" structures. The TFLN coupon is a piece of lithium niobate thin film with a tether system to suspend and hold the coupon in place. The suspended TFLN coupon is quickly picked up by an elastomeric stamp (PDMS stamp). Through a translation stage, the coupon is moved to the top of the designated position for printing over the target chip. The stamp is driving down towards the target chip and through slow retraction, the coupon is released from the stamp and transferred to the target chip. As discussed above, micro-transfer printing separates the processing of the source chip and target chip by confining the complex processing to the fabrication of the source chip. The preparation of the target chip starts with a 220 nm SOI chip. Two steps of E-beam lithography and ICP etching were conducted to define the full-etched waveguides and shallow-etched grating coupler. The preparation of the source chip refers to creation of a suspended "TFLN coupon" from the LNOI chip, as shown in Fig. 4(a). The coupon is fabricated in a rectangular shape with resist tethers on both sides. The fabrication process starts with an LNOI chip. Through photolithography and RIE etching, the coupon shape is patterned in the LN layer using an amorphous silicon hard mask, as shown in Fig. 4(b). Then, with another photolithography step and RIE etching, we patterned the release layer, as shown in Fig. 4(c). A third photolithography step is conducted to pattern the resist tethers. After a final selective release etching, the coupon is fully suspended, as shown in Fig. 4(d) and (e). Fig. 3 (c) shows the fabricated source chip. The interference fringes indicate the coupons in a suspended state.

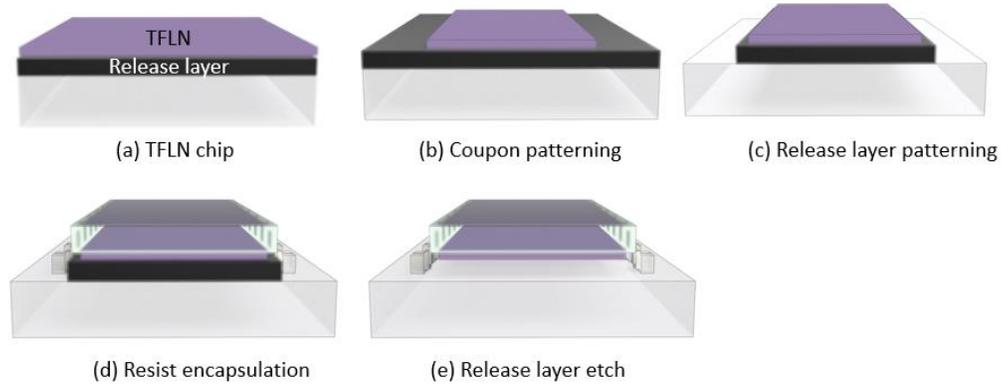

Fig. 4 Process steps for preparing "TFLN coupons" (a) Lithium niobate on insulator chip; (b) patterning of the coupon shape in the lithium niobate layer; (c) patterning of the release layer; (d) patterning of the resist encapsulation; (e) selective wet-etching.

Fig. 3 (c) shows an example of a printed coupon. One can observe that the coupon is uniformly attached to the silicon target and the tethers are breaking nicely. The printed chip was treated with acetone, IPA and an oxygen plasma clean process. After the micro-transfer printing, two metal lift-off steps were used to form the bottom electrodes and top electrodes. An E-beam lithography step was used to pattern the bottom electrodes, composed of 10 nm titanium and 700 nm gold. A 1 µm-thick $SiO_2$ layer is then deposited on top of the chip. Another E-beam lithography and BHF wet etching was conducted to form the vias. The top electrodes were then formed with a final E-beam lithography step and deposition of 10 nm titanium and 1500 nm gold. Fig. 5 (a) shows a microscope image of a part of the chip. Multiple thin-film lithium niobate coupons are printed on a compact area. Grating couplers are used to couple light in and out of the chip. Fig. 5 (b) shows the measured device. Fig. 5 (c), (d) and (e) are the false color SEM images of the device near the via of the contact region. We also made a FIB cut in this region. The SEM images of the cross-section are shown in Fig. 5 (d) and (e).

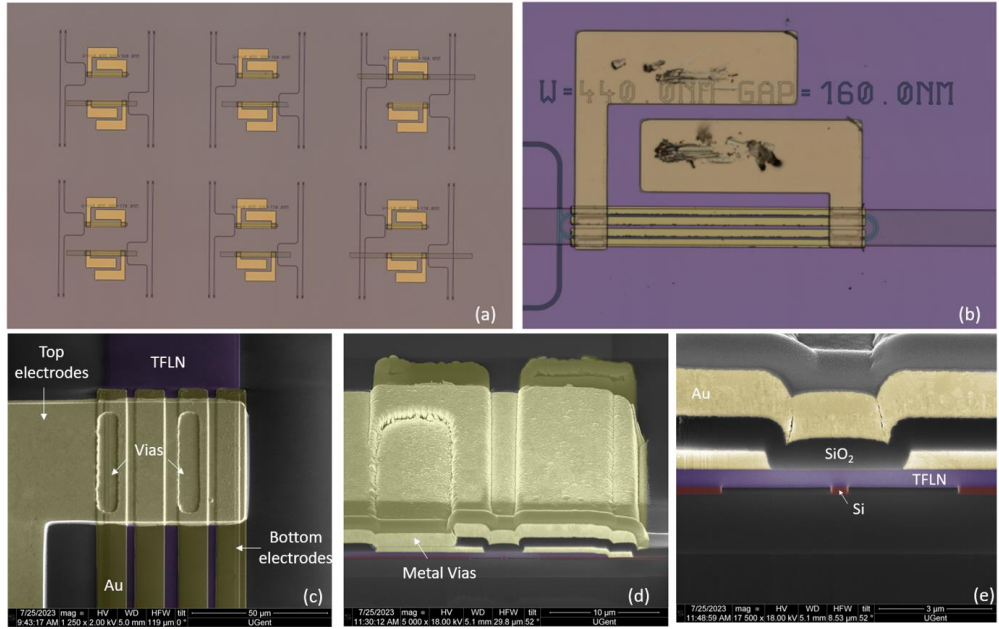

Fig. 5 (a) Microscope image of a section of the fabricated chip;(b) microscopic image of the micro-transfer printed thin-film lithium niobate ring modulator; (c) false-color SEM images of the vias region of the modulator (c) is the top view (d) and (e) are the cross-section.

## 3. Characterization

### 3.1 Transmission and Tuning efficiency

The device is characterized using a tunable laser and a power meter. The light is delivered to the chip by grating couplers. Fig. 6(a) is the measured transmission spectrum. The insertion loss of the device is around 1.5dB by normalizing to a 500nm-wide straight waveguide locally covered with a LN coupon of the same size. The extinction ratio of the resonances varies between 20dB and 37dB. The quality factor of the ring is around $1.118 \times 10^4$. Next, we measured the transmission over different applied voltages. We applied voltages in the range from −10V to 10V. The resonance of the ring modulator demonstrates a linear shift with a tuning efficiency of 3.89pm/V, as show in Fig. 6(b) and Fig. 6(c). The value is close to the simulation and the variations come from the fact that measurement was conducted at a longer wavelength.

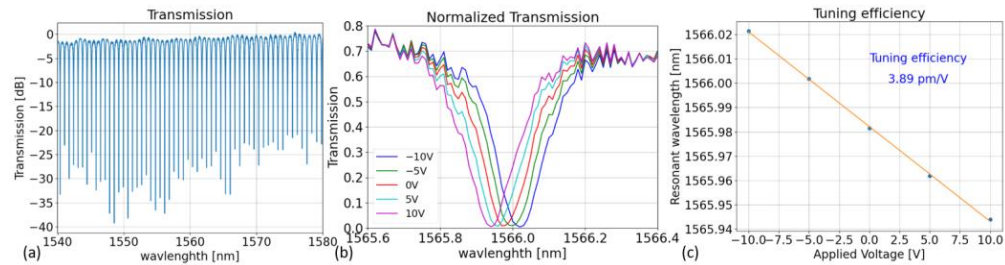

Fig. 6 Transmission measurement and tuning efficiency.

### 3.2 High-speed measurement

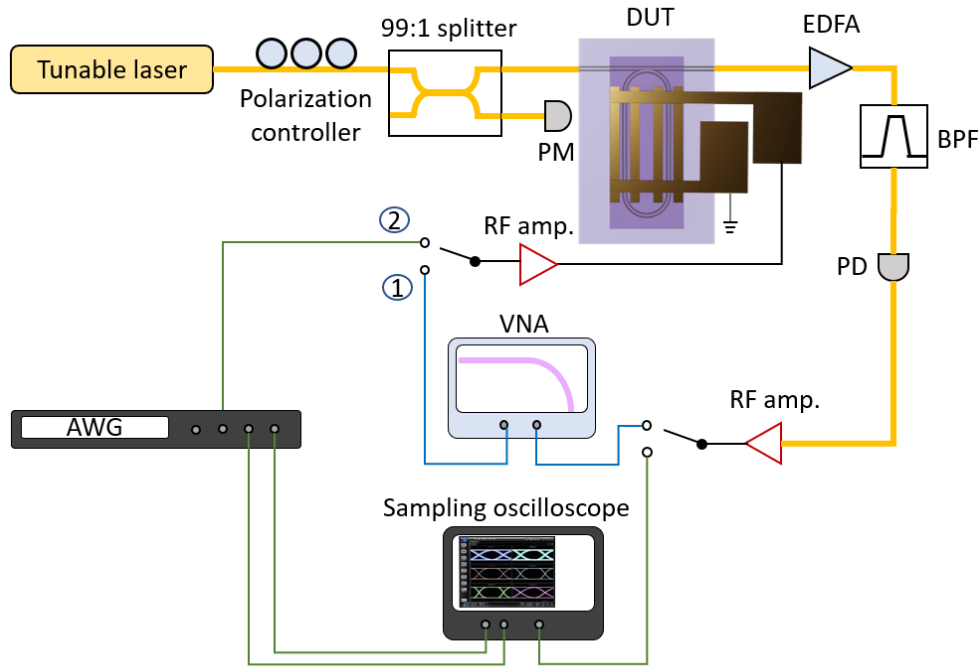

Fig. 7 Schematic of the measurement setup for bandwidth and high-speed data measurement. Light is coupled in and out of the chip through on chip grating coupler. A GS RF probe is used to electrically contact the circuit.

Fig. 7 shows the high-speed measurement setup. The source is a CW laser (Santec TSL510). The input light goes through a polarization controller and a power splitter allowing to send 1% of the signal to a power monitor (Newport 2936-R). The light is coupled to the chip through an on-chip grating coupler. The output from the chip is connected to an erbium doped fiber amplifier (EDFA, Keopsys CEFA-C-HG), goes through a bandpass filter and is then collected by a high-speed photodetector (FNINISSAR 50GHz photodetector XPDV21x0(RA)). The electro-optical bandwidth is characterized through electrical link 1. The RF signal generated by the vector network analyzer (VNA, Keysight N5247A PNA-X) is amplified by an RF amplifier (SHF S807C) and delivered to the chip by an RF probe (Picoprobe 50A-GS-100-DP). The output signal of the highspeed photodetector is amplified by an RF amplifier and collected by the VNA. Fig. 8(a) shows the measured electro-optic bandwidth of the ring modulator. The 3dB electro-optical bandwidth is measured to be 16GHz. The EO bandwidth is limited by cavity photon lifetime of the resonator(8.4ps), with a theoretical estimation of cavity photon lifetime limited EO bandwidth of 17GHz.

We also conducted a high-speed data measurement through electrical link 2. The time domain signal is generated by an Arbitrary Waveform Generator (Keysight M9502A AXle Chassis 8196A) and amplified to 2.6Vpp before sending to the chip. The applied signal is a non-return-to-zero pseudo-random binary sequence with length $2^7-1$ (PRBS7). The amplified signal from the PD is collected by a wide-bandwidth scope (Agilent DCA-X 86100D. Fig. 8 (c) shows the resulting eye diagrams. An open eye diagram is obtained up to 45Gbits$^{-1}$. We estimated the bit error rate (BER) of the system. According to Fig. 8 (b), BER is under $10^{-6}$ up to 36Gbits$^{-1}$ and $10^{-4}$ up to 40Gbits$^{-1}$.

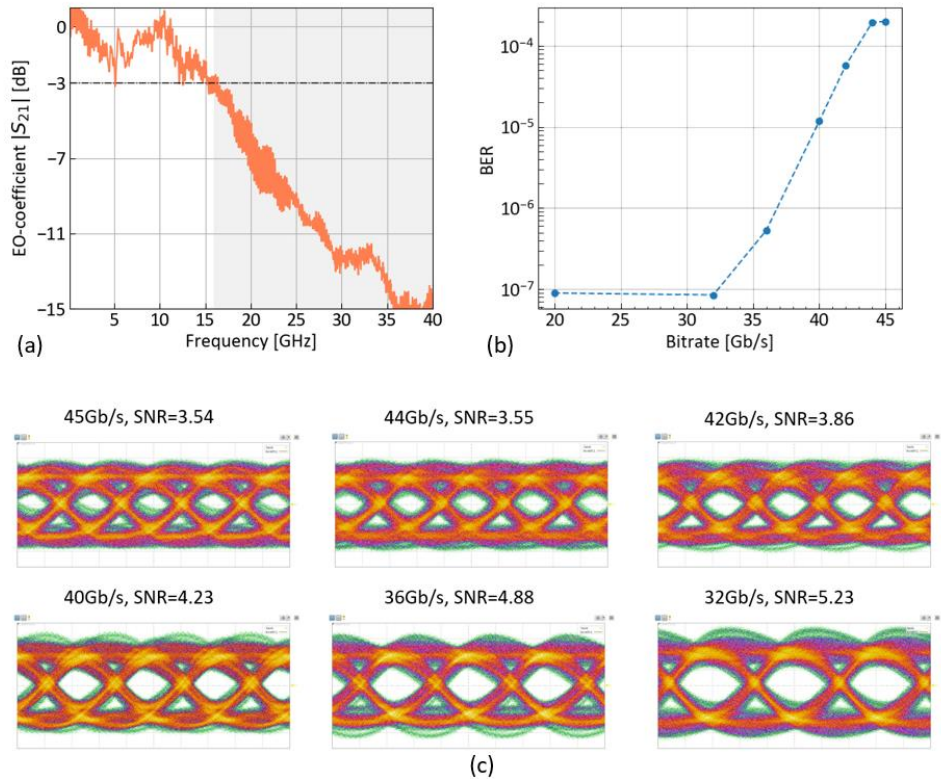

Fig. 8 The electro-optical bandwidth of the modulator.

## 4. Conclusion

In this paper, we demonstrated for the first time a micro-transfer-printed thin film lithium niobate (TFLN)-on-silicon ring modulator. Micro-transfer printing provides a heterogeneous solution to bring lithium niobate to CMOS-compatible platform with reduced fabrication complexity, increased integration density and cost-effectiveness. The micro-transfer-printed thin film lithium niobate (TFLN)-on-silicon ring modulator demonstrates high EO bandwidth (16GHz) and up to 45Gbits$^{-1}$ data transmission. This work paves the way towards dense integration of performant TFLN modulators with compact and scalable silicon circuity.

**Funding.** European Union's Horizon 2020 Research and Innovation Action SURQUID (899824).

**Acknowledgments.** Ying Tan thanks Muhammad Muneeb, Steven Verstuyft, Liesbet Van Landschoot for discussions on the fabrication process.

**Disclosures.**

**Data availability.** Data underlying the results presented in this paper are available in Dataset 1, Ref. [5].

**Supplemental document.** See Supplement 1 for supporting content.